\documentclass[aps,prl,showpacs]{revtex4}
\usepackage{amsmath}
\usepackage{graphicx}
\usepackage{bm}

\begin{document}

\title{Development of an approximate method for quantum optical models and their pseudo-Hermicity}
\date{\today}
\author{Ramazan Ko\c{c}}
\email{koc@gantep.edu.tr}
\affiliation{Department of Physics, Faculty of Engineering University of Gaziantep, 27310
Gaziantep, Turkey}

\begin{abstract}
An approximate method is suggested to obtain analytical
expressions for the eigenvalues and eigenfunctions of the some
quantum optical models. The method is based on the Lie-type
transformation of the Hamiltonians. In a particular case it is
demonstrated that $E\times \varepsilon $ Jahn-Teller Hamiltonian
can easily be solved within the framework of the suggested
approximation. The method presented here is conceptually simple
and can easily be extended to the other quantum optical models. We
also show that for a purely imaginary coupling the $E\times
\varepsilon $ Hamiltonian becomes non-Hermitian but $P\sigma
_{0}$-symmetric. Possible generalization of this approach is
outlined.
\end{abstract}

\pacs{03.65.Fd, 42.50.Ap}
\keywords{Algebraic Methods, Quantum Optical Models, Pseudo-Hermicity}
\maketitle

\section{Introduction}

It is well known that the rotating wave approximation (RWA) is a useful
method in determination of the eigenvalues and associated eigenfunctions of
the various quantum optical Hamiltonians. The approximation gives accurate
results when the frequency associated with the free evaluation of the system
is essentially bigger than the transmission frequencies induced by the
interaction between subsystem or external source. In quantum physics the
application of the RWA usually leads to symmetry breaking: the
representation space of the whole system is then divided into invariant
subspaces, which strongly simplifies the mathematical complexity of the
problem and usually provides the exact solution of the Hamiltonian.

The simplest model which describes a two-level atom interacting with a
single mode cavity field is the Jaynes-Cummings (JC) model \cite{jaynes}. A
considerable attention has been devoted to the interaction of a radiation
field with atoms since the paper of Dicke \cite{dicke}. Such system is
commonly termed as the Dicke model. In spite of its simplicity, the whole
spectrum of the Dicke Hamiltonian can not be obtained exactly and usually it
has been treated in the framework of RWA. Besides its solution with RWA, in
some papers an attempt is made to go beyond the RWA \cite{tur}. The
continual integration methods are based on variational principles. The
perturbative approach \cite{zaheer,zeng} leads to more complicated
mathematical treatments and the theory converges only for a certain
relationship between parameters of the Hamiltonian. In a more recent study,
Klimov and his co-workers \cite{klimov} have developed a general
perturbative approach to quantum optical models beyond the RWA, based on the
Lie-type transformation.

The Jahn Teller (JT) interaction \cite{jahn} is one of the most fascinating
phenomena in modern physics and chemistry, providing a general approach to
understanding the properties of molecules and crystals and their origins.
This phenomena has inspired of the most important recent scientific
discoveries, such as the concept of high temperature superconductivity. The
JT interaction is an example of electron-phonon coupling. Therefore it seems
that the RWA can be applied for solving the JT problems. Most of the JT
Hamiltonians are more complicated than the Dicke Hamiltonian. At present, a
few of them (i.e. $E\otimes \beta ,\ E\otimes \epsilon $ ) has been analyzed
in the framework of quasi-exactly solvable problem \cite{koc1,koc2} or
isolated exact solvability \cite{judd,longuet,reik,loorits,klenner,kus},
both provide finite number of exact eigenvalues and eigenfunctions in the
closed form. In this paper we devise a novel method for solving JT
Hamiltonians, as well as other quantum optical Hamiltonians in the framework
of RWA. It will be shown that the eigenvalues and the associated
eigenfunctions can be obtained in the closed form when the coupling constant
is smaller than the natural frequency of the oscillator.

The method described here includes a part of the motivation provided by the
existence of the connection between JT Hamiltonians and Dicke Model. Here we
concentrate our attention to the solution of the $E\otimes \epsilon $ \ JT\
Hamiltonian. Its solution has been treated previously by many authors \cite%
{judd,longuet,reik,kulak,lo,szopa}. We develop a new approximation method
which is based on the similarity transformation. The method introduced here
is the same as the RWA which has been usually used to solve Dicke
Hamiltonian. An interesting and somewhat simpler form of the JT Hamiltonian
is obtained by RWA.

Other purpose of this paper is to show that for some purely imaginary
couplings the $E\otimes \epsilon $ JT Hamiltonian becomes non-Hermitian but
its low-lying part of the spectrum is real. It will be shown that the
non-Hermitian Hamiltonian is not $PT$-invariant \cite%
{bender1,bender2,znojil,bagchi,ahmed} , but it is pseudo-Hermitian \cite%
{must1,must2,must3,bhabani,piju}.

In the following section, we shall demonstrate our procedure on the $%
E\otimes \epsilon $ JT Hamiltonian. We present a transformation procedure
and we obtain approximate form of the $E\otimes \epsilon $ JT Hamiltonian.
We show that Hamiltonian can be transformed in the form of the Dicke type
Hamiltonians. We also obtain explicit expressions for the eigenstates and
eigenvalues of the JT Hamiltonian. In section 3, we discuss the
pseudo-Hermicity of the Hamiltonian. Finally we summarize our results.

\section{Method and Summary of the Previous Results}

The well-known form of the $E\otimes \varepsilon $ JT Hamiltonian describing
a two-level fermionic subsystem coupled to two boson modes has obtained by
Reik\cite{reik} is given by%
\begin{equation}
H=\omega \left( a_{1}^{+}a_{1}+a_{2}^{+}a_{2}+1\right) +\omega _{0}\sigma
_{0}+\kappa \lbrack (a_{1}+a_{2}^{+})\sigma _{+}+(a_{1}^{+}+a_{2})\sigma
_{-}],  \label{1}
\end{equation}%
where $\omega _{0}$ is the level separation, $\omega $ is the frequency of
the oscillator and $\kappa $ is the coupling strength. The Pauli matrices $%
\sigma _{0,\pm }$ are given by%
\begin{equation}
\sigma _{+}=\left[ 
\begin{array}{cc}
0 & 1 \\ 
0 & 0%
\end{array}%
\right] ,\quad \sigma _{-}=\left[ 
\begin{array}{cc}
0 & 0 \\ 
1 & 0%
\end{array}%
\right] ,\;\sigma _{0}=\left[ 
\begin{array}{cc}
1 & 0 \\ 
0 & -1%
\end{array}%
\right] .  \label{2}
\end{equation}%
The annihilation and creation operators, $a_{i}\;$and$\;a_{i}^{+},$ satisfy
the usual commutation relations,%
\begin{equation}
\lbrack a_{i}^{+},a_{j}^{+}]=[a_{i},a_{j}]=0,\quad \lbrack
a_{i},a_{j}^{+}]=\delta _{ij}.  \label{eq:4}
\end{equation}%
The Hamiltonian (\ref{1}) can be solved in the framework of quasi-exactly
solvable problems\cite{koc2} or by using numerical diagonalization method%
\cite{tur}. In order to obtain rotating wave approximated form of the $%
E\otimes \varepsilon $ Hamiltonian, we use similarity transformation by
introducing the operator%
\begin{equation}
T=\frac{\kappa }{\omega +\omega _{0}}\left( \ \sigma _{+}a_{2}^{+}-\sigma
_{-}a_{2}\right) +\frac{\kappa }{\omega -\omega _{0}}\left( \ \sigma
_{-}a_{2}^{+}-\sigma _{+}a_{2}\right) ,  \label{3}
\end{equation}%
and imposing the condition \ $\left| \omega \pm \omega _{0}\right| \gg
\kappa ,$ which usually holds in the weak interaction, transformation of the
Hamiltonian (\ref{1}), yields that%
\begin{eqnarray}
\widetilde{H} &=&e^{T}He^{-T}\approx \omega \left(
a_{1}^{+}a_{1}+a_{2}^{+}a_{2}+1\right) +\omega _{0}\sigma _{0}+\kappa
\lbrack (a_{1}+a_{2})\sigma _{+}+(a_{1}^{+}+a_{2}^{+})\sigma _{-}]+  \notag
\\
&&\left[ \frac{\kappa ^{2}}{\omega +\omega _{0}}\left(
a_{1}^{+}a_{2}^{+}+a_{1}a_{2}\right) +\frac{\kappa ^{2}}{\omega -\omega _{0}}%
\left( a_{1}^{+}a_{2}+a_{1}a_{2}^{+}\right) \right] \sigma _{0}+  \notag \\
&&\frac{\omega \kappa ^{2}}{\omega ^{2}-\omega _{0}^{2}}\left(
a_{2}^{+2}+a_{2}^{2}+2a_{2}^{+}a_{2}\right) \sigma _{0}+  \label{4} \\
&&\frac{\kappa ^{2}\sigma _{+}\sigma _{-}}{\omega -\omega _{0}}-\frac{\kappa
^{2}\sigma _{-}\sigma _{+}}{\omega +\omega _{0}}+O\left( \frac{\kappa ^{3}}{%
\omega ^{2}-\omega _{0}^{2}}\right) .  \notag
\end{eqnarray}%
Since $\frac{\kappa ^{2}}{\omega \pm \omega _{0}}\ll 1$ is assumed to be a
small parameter, neglection of the last term confirms result;%
\begin{equation}
\widetilde{H}\approx \omega \left( a_{1}^{+}a_{1}+a_{2}^{+}a_{2}+1\right)
+\omega _{0}\sigma _{0}+\kappa \lbrack (a_{1}+a_{2})\sigma
_{+}+(a_{1}^{+}+a_{2}^{+})\sigma _{-}].  \label{5}
\end{equation}%
It is analytically solvable due to the neglect of \ the counter-rotating
terms, so called \ RWA. Now, we turn our attention to the solution of the
Hamiltonian (\ref{5}). The rotation of the bosons given by the following
operator%
\begin{equation}
U=\exp \left( \frac{\pi }{4}(a_{1}^{+}a_{2}-a_{2}^{+}a_{1}\right) 
\label{5a}
\end{equation}%
provides the expressions \ 
\begin{subequations}
\begin{eqnarray}
&&U(a_{1}+a_{2})U^{-1}=\sqrt{2}a_{1},\quad U(a_{1}^{+}+a_{2}^{+})U^{-1}=%
\sqrt{2}a_{1}^{+}  \notag \\
&&U(a_{1}^{+}a_{1}+a_{2}^{+}a_{2})U^{-1}=a_{1}^{+}a_{1}+a_{2}^{+}a_{2}
\label{5b}
\end{eqnarray}%
Under $U,$ the Hamiltonian becomes 
\end{subequations}
\begin{equation}
\widetilde{H}\approx \omega \left( a_{1}^{+}a_{1}+a_{2}^{+}a_{2}+1\right)
+\omega _{0}\sigma _{0}+\sqrt{2}\kappa \lbrack a_{1}\sigma
_{+}+a_{1}^{+}\sigma _{-}].  \label{5x}
\end{equation}%
The resultant Hamiltonian can easily be solved, because the matrix of the
Hamiltonian can be decomposed in infinite dependent $2\times 2$ blocks on
the subspaces $\left\{ \left| \uparrow ,n_{1}\right\rangle \left| \uparrow
,n_{2}\right\rangle ,\left| \downarrow ,n_{1}+1\right\rangle \left|
\downarrow ,n_{2}\right\rangle \right\} $, where $n_{1}$ and $n_{2}$ are the
number of photons.\ The eigenvalue problem can be written as 
\begin{table}[t]
\begin{tabular}{|c|c|c|c|c|}
\hline
& \multicolumn{2}{|c|}{Ground state} & \multicolumn{2}{|c|}{First excited
state} \\ \hline
$\kappa ^{2}$ & $E_{rwa}$ & $E_{exact}$ & $E_{rwa}$ & $E_{exact}$ \\ \hline
$0.1$ & 0.90455 & 0.90442 & 1.85982 & 1.82286 \\ \hline
$0.2$ & 0.81678 & 0.81595 & 1.73508 & 1.67515 \\ \hline
$0.3$ & 0.73508 & 0.73277 & 1.62159 & 1.54472 \\ \hline
$0.4$ & 0.65835 & 0.65371 & 1.51676 & 1.36373 \\ \hline
$0.5$ & 0.58578 & 0.57798 & 1.41886 & 1.31592 \\ \hline
$0.6$ & 0.51676 & 0.50498 & 1.32667 & 1.21248 \\ \hline
$0.7$ & 0.45080 & 0.43429 & 1.23931 & 1.11438 \\ \hline
$0.8$ & 0.38754 & 0.36557 & 1.15609 & 1.02070 \\ \hline
$0.9$ & 0.32667 & 0.29856 & 1.07646 & 0.93072 \\ \hline
\end{tabular}%
\caption{Ground-state and first excited-state energies of the $E\otimes 
\protect\varepsilon $ JT Hamiltonian.}
\label{tab:b}
\end{table}
\begin{equation}
\widetilde{H}\left| \psi \right\rangle =E\left| \psi \right\rangle 
\label{6}
\end{equation}%
where $\left| \psi \right\rangle $ is the two component \ eigenstate%
\begin{equation}
\left| \psi \right\rangle =\left( 
\begin{array}{l}
c_{1}\left| n_{1}\right\rangle \left| n_{2}\right\rangle  \\ 
c_{2}\left| n_{1}+1\right\rangle \left| n_{2}\right\rangle 
\end{array}%
\right) ,  \label{7}
\end{equation}%
where $c_{1}$and $c_{2\text{ }}$are normalization constant. Action of $%
\widetilde{H\text{ }}$on $\psi $ yields the following expressions 
\begin{subequations}
\begin{align}
\left( c_{1}\left( \omega \left( n_{1}+n_{2}+1\right) +\omega _{0}\right)
+c_{2}\sqrt{2}\kappa \sqrt{n_{1}+1}\right) \left| n_{1}\right\rangle \left|
n_{2}\right\rangle & =c_{1}E\left| n_{1}\right\rangle \left|
n_{2}\right\rangle   \label{8a} \\
\left( c_{2}\left( \omega \left( n_{1}+n_{2}+2\right) -\omega _{0}\right)
+c_{1}\sqrt{2}\kappa \sqrt{n_{1}+1}\right) \left| n_{1}+1\right\rangle
\left| n_{2}\right\rangle & =c_{2}E\left| n_{1}+1\right\rangle \left|
n_{2}\right\rangle .  \label{8b}
\end{align}%
Eliminating $c_{1}$ and $c_{2\text{ }}$between (\ref{8a}) and (\ref{8b}) and
solving the resultant equation \ for $E$, we obtain 
\end{subequations}
\begin{equation}
E=\left( j+1\right) \omega \pm \frac{1}{2}\sqrt{8\kappa ^{2}(n+1)+\left(
\omega -2\omega _{0}\right) ^{2}}.  \label{9}
\end{equation}%
where $j=n_{1}+n_{2}$ total number of bosons and $n=0,1,2,\cdots ,2j.$ The
eigenstates can be easily written by using boson operators, acting on a
vacuum state$\left| 0\right\rangle ;$%
\begin{equation}
\left| \psi \right\rangle =\left[ c_{1}a_{2}^{j-n}a_{1}^{+n}\left|
0\right\rangle ,c_{2}a_{2}^{j-n}a_{1}^{+n+1}\left| 0\right\rangle \right]
^{T}.  \label{10}
\end{equation}%
We conclude that in weak coupling limit the oscillators does not coupled to
each other and each of them oscillates with their own frequencies. We have
proven that when the interaction between $E$ ion and $\varepsilon $-modes
are weak then $E\otimes \varepsilon $ JT Hamiltonian can be reduced to the
JC model. Our formalism provides a solution of the problem which allows us
to discuss the JT effects in the Dicke model.

The accuracy of the approximate eigenvalues can be checked by means of the
(quasi) exact solution of the $E\otimes \varepsilon $ JT Hamiltonian. The
material parameters are chosen to be $\omega =1$ and $\omega _{0}=0$. The
results are tabulated in Table 1.

The results of our study show that the eigenvalues and eigenstates \ of the $%
E\otimes \varepsilon $ JT\ Hamiltonian can be approximately described when
the frequency $\omega $ of the oscillation larger than the interaction
constant.

\section{Non-Hermitian interaction}

It has been shown that for some purely imaginary couplings constant $\kappa ,
$ the low-lying part of the $E\otimes \varepsilon $ JT Hamiltonian is real,
although the Hamiltonian is non-Hermitian. Let us consider the Hamiltonian (%
\ref{5x}) with the imaginary coupling $\kappa =i\gamma :$%
\begin{equation}
h=\omega \left( a_{1}^{+}a_{1}+a_{2}^{+}a_{2}+1\right) +\omega _{0}\sigma
_{0}+i\sqrt{2}\gamma \lbrack a_{1}\sigma _{+}+a_{1}^{+}\sigma _{-}].
\label{11}
\end{equation}%
This Hamiltonian is not Hermitian as,%
\begin{equation}
h^{\dagger }=\omega \left( a_{1}^{+}a_{1}+a_{2}^{+}a_{2}+1\right) +\omega
_{0}\sigma _{0}-i\sqrt{2}\gamma \lbrack a_{1}\sigma _{+}+a_{1}^{+}\sigma
_{-}]\neq h.  \label{12}
\end{equation}%
Under the parity transformation, the Pauli matrices become invariant but
both the creation and annihilation operators change sign. The time reversal
operator for this Hamiltonian is $T=-i\sigma _{y}K$ where $K$ is complex
conjugation operator. The time reversal operator changes the sign of the
Pauli matrices and boson operators. \ It is easy to see that the Hamiltonian
(\ref{11}) is not $PT$-symmetric%
\begin{equation}
(PT)h(PT)^{-1}=\omega \left( a_{1}^{+}a_{1}+a_{2}^{+}a_{2}+1\right) -\omega
_{0}\sigma _{0}+i\sqrt{2}\gamma \lbrack a_{1}\sigma _{+}+a_{1}^{+}\sigma
_{-}]\neq h.  \label{13}
\end{equation}%
The Hamiltonian is not $PT$-symmetric but it gives real spectrum.
Mustafazadeh \cite{must1,must2,must3} has shown that the reality of the
spectrum of non-Hermitian Hamiltonian is due to pseudo-Hermicity properties
of the Hamiltonian. A Hamiltonian is called $\eta $-pseudo-Hermitian if it
satisfies the following relation%
\begin{equation}
\eta h\eta ^{-1}=h^{\dagger },
\end{equation}%
where $\eta $ is a linear Hermitian operator. The Hamiltonian $h$ and its
adjoint $h^{\dagger }$ can be related to each others by the operator $\sigma
_{0}$ and using the relation $\sigma _{0}\sigma _{\pm }\sigma
_{0}^{-1}=-\sigma _{\pm }:$ 
\begin{equation}
\sigma _{0}h\sigma _{0}^{-1}=h^{\dagger }.
\end{equation}%
Then the Hamiltonian (\ref{11}) is $\sigma _{0}$-pseudo-Hermitian. Our
Hamiltonian is also pseudo-Hermitian with respect to the parity operator. As
it is shown \cite{must1} that if a Hamiltonian is pseudo-Hermitian under two
different operators, $\eta _{1},$ $\eta _{2}$ then the system is symmetric
under the transformation generated by $\eta _{1}\eta _{2}^{-1}.$ Therefore
our Hamiltonian is invariant under the symmetry generated by the combined
operator, $P\sigma _{0}:$%
\begin{equation}
\left[ H,P\sigma _{0}\right] =0.
\end{equation}

\section{Conclusion}

The aim of the this paper was to illustrate how the $E\otimes \varepsilon $
JT Hamiltonians can be solved by developing a transformation procedure. It
has been found an approximate form of the $E\otimes \varepsilon $ JT
Hamiltonian in the framework of the RWA. The resultant Hamiltonian can be
solved analytically and its eigenvalues can be obtained in the closed form.
We have shown that in the weak coupling limit the JT models may be
recognized as the Dicke model. We have shown that when the coupling constant
is imaginary the Hamiltonian is non-Hermitian but $P\sigma _{0}$-symmetric.
We also hope to extend the method to the other JT and quantum optical
systems.

\bigskip 

\end{document}